# Reconstruction of complete interval tournaments

Antal Iványi
Eötvös Loránd University,
Department of Computer Algebra
1117 Budapest, Pázmány Péter sétány 1/C.
email: `tony@compalg.inf.elte.hu`

**Abstract.** Let $a$, $b$ and $n$ be nonnegative integers ($b \geq a$, $b > 0$, $n \geq 1$), $\mathcal{G}_n(a, b)$ be a multigraph on $n$ vertices in which any pair of vertices is connected with at least $a$ and at most $b$ edges and $\mathbf{v} = (v_1, v_2, \ldots, v_n)$ be a vector containing $n$ nonnegative integers. We give a necessary and sufficient condition for the existence of such orientation of the edges of $\mathcal{G}_n(a, b)$, that the resulted out-degree vector equals to $\mathbf{v}$. We describe a reconstruction algorithm. In worst case checking of $\mathbf{v}$ requires $\Theta(n)$ time and the reconstruction algorithm works in $O(bn^3)$ time. Theorems of H. G. Landau (1953) and J. W. Moon (1963) on the score sequences of tournaments are special cases $b = a = 1$ resp. $b = a \geq 1$ of our result.

## 1 Introduction

Ranking of objects is a typical practical problem. One of the popular ranking methods is the pairwise comparison of the objects. If the result of a comparison is expressed by dividing points between the corresponding objects, then directed graphs serve as natural tools to represent the results: vertices correspond to the objects, arcs to the points and out-degrees serve as basis for ranking. Another natural tool to represent the results is a point table.

In this paper the terminology of D. E. Knuth [9] and the pseudocode of T. H. Cormen and his coauthors [2] are used.

Let $a$, $b$ and $n$ be nonnegative integers ($b \geq a$, $n \geq 1$), $\mathcal{T}_n(a,b)$ be a directed multigraph on $n$ vertices in which any pair of vertices is connected with at least $a$ and at most $b$ arcs. Then $\mathcal{T}_n(a,b)$ is called **interval** or $(a,b)$**-tournament**, its vertices are called **players,** the out-degree sequence $\mathbf{v} = (v_1, v_2, \ldots, v_n)$ is called **score vector** and the comparisons are called **matches**.

For the simplicity we suppose that $v_1 \leq v_2 \leq \cdots \leq v_n$. The increasingly ordered score vector is called **score sequence** and is denoted by $\mathbf{s} = (s_1, s_2, \ldots, s_n)$.

If any integer partition of the points is permitted, then the tournament is **complete**, otherwise **incomplete** [7].

If $a = b \geq 1$, then we get multitournaments $\mathcal{T}_n(a)$ and if $a = b = 1$, then we get the well-known concept of tournaments $\mathcal{T}_n$.

In 1953 H. G. Landau [10] proved the following popular theorem. About ten proofs are summarised by K. B. Reid [14] and two recent ones are due to J. Griggs and K. B. Reid [4], resp. to K. B. Reid and C. Q. Zhang [15]. Pirzada, Shah and Naikoo investigated similar problems [13]. Several exercises on tournaments can be found in the recent book of D. E. Knuth [8].

**Theorem 1** *A sequence* $(s_1, s_2, \ldots, s_n)$ *satisfying* $0 \leq s_1 \leq s_2 \leq \ldots \leq s_n$ *is the score sequence of some tournament* $\mathcal{T}_n(1)$ *if and only if*

$$\sum_{i=1}^{k} s_i \geq B_k, \quad 1 \leq k \leq n, \tag{1}$$

*with equality when* $k = n$.

In 1963 J. W. Moon in [11] proved the following generalisation of the Landau's theorem.

**Theorem 2** *A sequence* $(s_1, s_2, \ldots, s_n)$ *satisfying* $0 \leq s_1 \leq s_2 \leq \cdots \leq s_n$ *is the score sequence of some a-tournament* $\mathcal{T}_n(a)$ *if and only if*

$$\sum_{i=1}^{k} s_i \geq aB_k, \ 1 \leq k \leq n, \tag{2}$$

*with equality when* $k = n$.

Figure 1 shows the point table of a tournament $\mathcal{T}_6(2, 10)$. The score sequence of this tournament is $\mathbf{s} = (9,9,19,20,32,34)$.

| Player/Player | $\mathcal{P}_1$ | $\mathcal{P}_2$ | $\mathcal{P}_3$ | $\mathcal{P}_4$ | $\mathcal{P}_5$ | $\mathcal{P}_6$ | Score |
|---|---|---|---|---|---|---|---|
| $\mathcal{P}_1$ | — | 1 | 5 | 1 | 1 | 1 | 9 |
| $\mathcal{P}_2$ | 1 | — | 4 | 2 | 0 | 2 | 9 |
| $\mathcal{P}_3$ | 3 | 3 | — | 5 | 4 | 4 | 19 |
| $\mathcal{P}_4$ | 8 | 2 | 5 | — | 2 | 3 | 20 |
| $\mathcal{P}_5$ | 9 | 9 | 5 | 7 | — | 2 | 32 |
| $\mathcal{P}_6$ | 8 | 7 | 5 | 6 | 8 | — | 34 |

Figure 1: The results of the matches of six players.

We wish to decide whether there exist tournaments with a given score sequence and if yes, then we wish to reconstruct one of them.

Our problems can be formulated also as follows [3]. Let $\mathcal{G}_n$ be a multigraph in which the number of connecting edges lies between $a$ and $b$ for any pair of vertices. Design effective algorithms to decide whether there exist an orientation of the edges guaranteeing a prescribed out-degree sequence and to reconstruct a corresponding digraph.

We remark that Gyárfás et al. [5] and Brualdi [1] published quick algorithms for 1-tournaments.

Also it is worth to remark that many enumeration type results are known. In connection with classical tournaments it is known due to P. Tetali [16] that only a few score sequences permit the reconstruction in a unique way: typical is the large number of nonisomorph reconstructions. G. Péchy and L. Szűcs [12] proposed a parallel algorithm for generation of all possible score sequences of the 1-tournaments of $n$ players.

The aim of this paper is to solve the decision and reconstruction problems [6] for complete $(a, b)$-tournaments.

## 2 Necessary conditions for $(a, b)$-tournaments

It is easy too see the following necessary condition, where $B_n$ is the binomial coefficient $n$ over 2 for $n = 1, 2, \ldots$.

**Lemma 1** *If* $(s_1, s_2, \ldots, s_n)$ *is the score sequence of some* $(a, b)$*-tournament* $T_n(a, b)$*, then*

$$\sum_{i=1}^{k} s_i \geq aB_k \quad (1 \leq k \leq n) \tag{3}$$

*and*

$$\sum_{i=1}^{n} s_i \leq bB_n. \qquad (4)$$

If $a = 2$ and $b = 10$, then the sequence $\mathbf{s} = (1, 1, 21)$ shows that the requirements of Lemma 1 are not sufficient. Since $\mathcal{P}_1$ and $\mathcal{P}_2$ divided only 2 points, they lost at least 8 points and so the sum of the scores can be at most 22 instead of $b\mathcal{B}_3 = 30$. This remark can be extended to a general condition.

We define a loss function $L_k$ ($k = 0, 1, 2, \ldots, n$) by the following recursion: $L_0 = 0$ and if $1 \leq k \leq n$, then

$$L_k = \max\left(L_{k-1}, \ bB_k - \sum_{i=1}^{k} s_i\right). \qquad (5)$$

Now $L_k$ gives a lower bound for the number of lost points in the matches among the players $\mathcal{P}_1$, $\mathcal{P}_2$, ..., $\mathcal{P}_k$ (not always the exact value since the players $\mathcal{P}_1$, $\mathcal{P}_2$, ..., $\mathcal{P}_k$ could win points against $\mathcal{P}_{k+1}, \ldots, \mathcal{P}_n$).

**Lemma 2** *If $(s_1, s_2, \ldots, s_n)$ is the score sequence of some $(a, b)$-tournament $\mathcal{T}_n(a, b)$, then*

$$\sum_{i=1}^{k} s_i + (n - k)s_k \leq bB_n - L_k \quad (1 \leq k \leq n). \qquad (6)$$

**Proof.** The member $(n - k)s_k$ of the left side is due to the monotonicity of $\mathbf{s}$. The loss function $L_k$ takes into account the lost points of the matches among the players $\mathcal{P}_1, \ldots, \mathcal{P}_k$. ∎

These lemmas imply the following assertion.

**Lemma 3** *If $(s_1, s_2, \ldots, s_n)$ is the score sequence of some $(a, b)$-tournament $\mathcal{T}_n(a, b)$, then*

$$a\mathcal{B}_k \leq \sum_{i=1}^{k} s_i \leq bB_k - L_k - (n - k)s_k \quad (1 \leq k \leq n). \qquad (7)$$

**Proof.** (7) is an algebraic consequence of (3) and (6). ∎

## 3  Definition of the algorithms

We describe the proposed new algorithms in words, by examples and by the pseudocode used in [2].

Algorithm SCORECHECK uses Lemma 3. Algorithm SCORESLICING is an extended version of Ryser's construction method [14], and algorithm MAIN organises the work of SCORESLICING.

At first let's consider the small tournament $\mathcal{T}_3(2,10)$ whose point table is shown in Figure 2. The score sequence of this tournament is $\mathbf{s} = (3,4,5)$.

| Player/Player | $\mathcal{P}_1$ | $\mathcal{P}_2$ | $\mathcal{P}_3$ | Score |
|---|---|---|---|---|
| $\mathcal{P}_1$ | — | 3 | 0 | 3 |
| $\mathcal{P}_2$ | 0 | — | 4 | 4 |
| $\mathcal{P}_3$ | 4 | 1 | — | 5 |

Figure 2: The results of the matches of three players.

According to (5) we have $L_0 = 0$, $L_1 = 0$, $L_2 = bB_2 - S_2 = 3$, and $L_3 = bB_3 - S_3 = 18$. The requirements of Lemma 3 are $aB_1 = 0 \leq S_1 \leq bB_3 - 2s_1 = 24$, $aB_2 = 2 \leq S_2 \leq bB_3 - L_2 - s_2 = 23$ and $aB_3 = 6 \leq S_3 \leq bB_3 - L_3 = 12$. These inequalities hold.

Let's try to construct a possible point table. The number of points of $\mathcal{P}_i$ against $\mathcal{P}_j$ is denoted by $r_{i,j}$ ($1 \leq i, j \leq n$). Provisionally we suppose $r_{i,j} = b = 10$, if $j > i$, and $r_{i,j} = 0$ otherwise (in the main diagonal of the table $r_{ij} = 0$ is represented by –).

We begin with the possible results of the player $\mathcal{P}_3$ having the largest number of points. We fix such results for $\mathcal{P}_3$ that after removing of its results from the point table the score sequence $(s'_1, s'_2)$ of the remaining players is monotone and satisfies (7).

$\mathcal{P}_3$ has only $s_3 = 5$ points instead of the possible maximum $(n-1)b = 20$, so $M_3 = 20 - 5 = 15$ points are missing. These points are win by other players or are lost. At first we determine the points win by other players, then the points lost by $\mathcal{P}_3$.

How many is the maximal permitted value of $r_{2,3}$? Since we investigate a (2,10)-tournament, $r_{2,3} \leq b = 10$. $\mathcal{P}_1$ and $\mathcal{P}_2$ play a match where they together have to win at least $a = 2$ points, therefore they can win against $\mathcal{P}_3$ at most $A_2 = s_1 + s_2 - aB_1 = 5$ additional points, so $r_{2,3} \leq A_2 = 5$. A natural requirement is $r_{2,3} \leq s_2 = 4$. The monotonicity requires $r_{2,3} \leq s_2 - s_1 = 1$.

The strongest requirement is $r_{2,3} \leq 1$, therefore let $r_{2,3} = 1$. So we founded place for 1 point from the 15 missing points of $\mathcal{P}_3$, the score sequence of the modified $\mathcal{T}_2$ is $(3,3)$, $\mathcal{P}_1$ and $\mathcal{P}_2$ have $A_2' = s_1' + s_2' - aB_1 = 4$ additional points and $M_3' = 14$.

We divide these additional points between $\mathcal{P}_1$ and $\mathcal{P}_2$ and get $r_{2,3}'' = 1+2 = 3$, $r_{1,3}'' = 0+2 = 2$ and $M_3'' = 10$. These numbers imply $r_{3,2}' = b - r_{2,3}'' = 7$ and $r_{3,1}' = b - r_{1,3}'' = 8$. Since $A_2 = 0$, that is $\mathcal{P}_1$ and $\mathcal{P}_2$ have no further additional points, they can not win further points from $\mathcal{P}_3$. $\mathcal{P}_3$ lost $r_{2,3} + r_{1,3} = 3+2 = 5$ points, so we found 5 of the missing $M_3 = 15$ points. Now we determine $r_{3,2}'$ trying to decrease $M_3''$ as possible. Since $r_{2,3}''$ is large enough to guarantee $r_{2,3} + r_{3,2} \geq a$ and $M_3'' = 10$ is also large enough, let $r_{3,2}' = 0$ implying $M_3''' = 10 - 7 = 3$. The next step is to fix $r_{3,1}'' = r_{3,1}' - M_3''' = 8 - 3 = 5$. Now $\mathcal{P}_3$ has the obligatory 5 points, and $\mathcal{P}_1$ needs further $s_1'' = s_1' - r_{1,3}'' = 1$ point, and $\mathcal{P}_2$ needs further $s_2'' = s_2' - r_{2,3}'' = 1$ point. So we can remove $\mathcal{P}_3$ receiving a tournament $\mathcal{T}_2(2, 10)$ with a score sequence $\mathbf{s}'' = (1,1)$ and we can finish the construction setting $r_{1,2} = 1$ and $r_{2,1} = 1$.

The following Figure 3 shows the reconstructed tournament.

| Player/Player | $\mathcal{P}_1$ | $\mathcal{P}_2$ | $\mathcal{P}_3$ | Score |
|---|---|---|---|---|
| $\mathcal{P}_1$ | — | 1 | 2 | 3 |
| $\mathcal{P}_2$ | 1 | — | 3 | 4 |
| $\mathcal{P}_3$ | 5 | 0 | — | 5 |

Figure 3: The reconstructed results of the matches of three players.

In this simple example we can answer the question: how many possible reconstructions are possible? Since $r_{1,2}$ and $s_1$ determine $r_{1,3}$, $r_{2,1}$ and $s_2$ determine $r_{2,3}$, $r_{3,1}$ and $s_3$ determine $r_{3,2}$, we have at most $(s_1 + 1) \times (s_2 + 1) \times (s_3 + 1) = 120$ reconstructions.

The exact value of the number of the possible reconstructions is smaller. For example the permitted values of $r_{1,2}$ are 0, 1, 2, and 3. But if $r_{1,2} = 2$, then $r_{1,3} = s_1 - r_{1,2} = 1$. Now $r_{3,1} + r_{1,3} \geq a = 2$ and $r_{3,1} \leq s_5$ allow only 1, 2, 3, 4 and 5 for $s_{3,1}$, that is there are only 5 possibilities instead of 6.

### 3.1 Definition of the checking algorithm

*Input.* $a$ and $b$: minimal and maximal number of points divided after each match;
$n =:$ the number of players ($n \geq 2$);

$\mathbf{s} = (s_1, s_2, \ldots, s_n)$: a nondecreasing sequence of integers.
  *Output.* One of the following messages:
$i$"-th score is too small";
$i$"-th score is too large";
"the sequence satisfies both necessary conditions";
$\mathbf{B} = (B_0, B_1, \ldots, B_n)$: the sequence of the binomial coefficients;
$\mathbf{L} = (L_0, L_1, \ldots, L_n)$: the sequence of the values of the loss function;
$\mathbf{S} = (S_0, S_1, \ldots, S_n)$: the sequence of the sums of the $i$ smallest scores.
  *Working variables.* $i$: cycle variable.

SCORECHECK($n, a, b, \mathbf{B}, \mathbf{L}, \mathbf{s}, \mathbf{S}$)
**01** $L_0 \leftarrow 0$
**02** $S_0 \leftarrow 0$
**03** $B_0 \leftarrow 0$
**04 for** $i \leftarrow 1$ **to** $n$
**05**     **do** $S_i \leftarrow S_{i-1} + s_i$
**06**         $B_i \leftarrow B_{i-1} + i - 1$
**07**         $L_i \leftarrow \max(L_{i-1}, bB_i - S_i)$
**08**         **if** $S_i < aB_i$
**09**             **then return** $i$"-th score is too small"
**10**         **if** $S_i > bB_n - L_i - s_i(n-i)$
**11**             **then return** $i$"-th score is too large"
**12 return** "the sequence satisfies both necessary conditions"

Figure 1 shows the point table of a tournament of 6 players. In this case the score sequence is $\mathbf{s} = (9, 9, 19, 20, 32, 34)$, $L_0 = 0$, $L_1 = 0$, $L_2 = 0$, $L_3 = 0, L_4 = 3, L_5 = 11$, and $L_6 = 27$. The requirements of (7) are fulfilled: $0 \leq S_1 = 9 \leq 105$, $2 \leq S_2 = 18 \leq 114$, $6 \leq S_3 = 37 \leq 93$, $12 \leq S_4 = 57 \leq 107$, $20 \leq S_5 = 89 \leq 107$, $30 \leq S_6 = 123 \leq 123$. Therefore the conditions in lines 08 and 10 of this program never hold, so the algorithm returns the message of line 12.

### 3.1.1 Complexity analysis of the checking algorithm

The running time of SCORECHECK is $\Theta(n)$ in worst case.

  For incorrect sequences the running time of SCORECHECK can be small. For example if $s_1 = s_2 = (n-1)b$ or $a > 0$ and $s_1 = s_2 = 0$, then the running time is $O(1)$.

  We remark that adding a linear time sorting algorithm [2] SCORECHECK can be extended for score vectors too (saving the linear running time).

The memory requirement of SCORECHECK is $\Theta(n)$. If the stepwise input of the scores is permitted, then we can implement this algorithm using only $O(1)$ memory.

## 3.2 Definition of the main algorithm

The work of the slicing program is managed by the following program MAIN.

*Input.* $a$ and $b$: minimal and maximal number of points divided after each match;
$\mathbf{B} = B_0, B_1, \ldots, B_n)$: the sequence of the binomial coefficients;
$\mathbf{L} = (L_0, L_1, \ldots, L_n)$: the values of the loss function;
$n$: the number of players ($n \geq 2$);
$\mathbf{s} = (s_1, s_2, \ldots, s_n)$: a nondecreasing sequence of integers satisfying (7);
$\mathbf{S} = (S_1, S_2, \ldots, S_n)$: the sums of the scores.

*Output.* $R = [r_{i,j}]_{n \times n}$: point table of the reconstructed tournament $\mathcal{T}_n(a, b)$.

*Working variables.* $g$, $i$, $k$: cycle variables;
$\mathbf{p} = (p_1, p_2, \ldots, p_n)$: a provisional score sequence;
$\mathbf{p}_k = (p_1, p_2, \ldots, p_k)$ ($k = 1, 2, \ldots, n$): prefixes of the provisional score sequence $\mathbf{p}$;
$\mathbf{q} = (q_1, q_2, \ldots, q_{k-1}) = (r_{1,k}, r_{2,k}, \ldots, r_{k-1,k})$;
$\mathbf{r} = (r_1, r_2, \ldots, r_{k-1}) = (r_{k,1}, r_{k,2}, \ldots, r_{k,k-1})$.

During the reconstruction process we have to take into account the following bounds:

$$a \leq r_{i,j} + r_{j,i} \leq b \quad (1 \leq i, \ j \leq n, i \neq j); \tag{8}$$

$$\text{modified scores have to satisfy (7);} \tag{9}$$

$$r_{i,j} \leq p_i \ (1 \leq i, \ j \leq n, i \neq j); \tag{10}$$

the monotonicity $p_1 \leq p_2 \leq \ldots \leq p_k$ has to be saved $(1 \leq k \leq n)$. $\tag{11}$

MAIN$(a, b, n, \mathbf{B}, \mathbf{L}, \mathbf{p}, \mathcal{R})$

**01 for** $i \leftarrow 1$ **to** $n$
**02**     **do** $\mathcal{R}_{i,i} \leftarrow 0$
**03**         $p_i \leftarrow s_i$

```
04 if n ≥ 3
05    then for k ← n downto 3
06            do SCORESLICING(a, b,B,L,k,p_{k-1},p_k)
07               for g ← 1 to k − 1
08                  do R_{g,k} ← q_g
09                     R_{k,g} ← r_g
10 r_{1,2} ← ⌊(p_1 + p_2)/2⌋
11 r_{2,1} ← ⌈(p_1 + p_2)/2⌉
12 return R
```

### 3.3 Definition of the slicing algorithm

The key part of the reconstruction is the following algorithm SCORESLICING.

*Input.* $a$, $b$: minimal and maximal number of points divided after each match;
$\mathbf{B} = (B_1, B_2, \ldots, B_n)$: the sequence of the binomial coefficients;
$\mathbf{L} = (L_1, L_2, \ldots, L_k)$: the values of the loss function;
$k$: the number of the actually investigated players ($k > 2$);
$\mathbf{p}_k = (p_1, p_2, \ldots, p_k)$: provisional score sequence;
$\mathbf{s} = (s_1, s_2, \ldots, s_k)$: a nondecreasing sequence of integers satisfying (7);
$\mathbf{S} = (S_1, S_2, \ldots, S_k)$: the sums of the scores.

*Output:* $\mathbf{p}_{k-1} = (p_1, p_2, \ldots, p_{k-1})$: a provisional score sequence;
$\mathbf{q} = (q_1, q_2, \ldots, q_{k-1}) = (r_{1,k}, r_{2,k}, \ldots, r_{k-1,k})$;
$\mathbf{r} = (r_1, r_2, \ldots, r_{k-1}) = (r_{k,1}, r_{k,2}, \ldots, r_{k,k-1})$.

*Working variables.* $\mathbf{A} = (A_1, A_2, \ldots, A_n)$ the number of the additional points;
$d$: difference of the maximal increasable scores and the following largest score;
$e$: number of sliced points per player;
$f$: frequency of the number of maximal values among the scores $p_1, p_2, \ldots, p_{k-1}$;
$g$, $h$, $i$: cycle variables;
$m$: maximal amount of sliceable points;
$M$: missing points: the difference of the number of actual points and the number of maximal possible points of $\mathcal{P}_k$;
$p_0$: number of points of the hypothetical "negative player" $\mathcal{P}_0$ used in line 15;
$\mathbf{P} = (P_1, P_2, \ldots, P_n)$: the sums of the provisional scores;
$x$: the maximal index $i$ with $i < k$ and $r_{i,k} < b$.

SCORESLICING($a, b, \mathbf{B}, \mathbf{L}, n, \mathbf{p}_{k-1}, \mathbf{p}_k$)

**01** $p_0 \leftarrow 0$
**02** $P_0 \leftarrow 0$
**03 for** $i \leftarrow 1$ **to** $k - 1$
**04**     **do** $P_i \leftarrow P_{i-1} + p_i$
**05**         $A_i \leftarrow P_i - aB_i$
**06 for** $g \leftarrow 1$ **to** $k - 1$
**07**     **do** $r_{g,k} \leftarrow 0$;
**08**         $r_{k,g} \leftarrow b$;
**09** $M \leftarrow (k-1)b - p_k$
**10 while** $M > 0$ **and** $A_{k-1} > 0$
**11**     **do** $x \leftarrow k - 1$
**12**         **while** $r_{x,k} = b$
**13**             **do** $x \leftarrow x - 1$
**14**     $f \leftarrow 1$
**15**     **while** $p_{x-f+1} = p_{x-f}$
**16**         **do** $f = f + 1$
**17**     $d \leftarrow p_{x-f+1} - p_{x-f}$
**18**     $m \leftarrow \min(b, d, \lceil A_x/f \rceil, \lceil M/f \rceil)$
**19**     **for** $g \leftarrow f$ **downto** $1$
**20**         **do** $y \leftarrow \min(b - r_{x+1-g,k}, m, M, A_{x+1-g}, p_{x+1-g})$
**21**             $r_{x+1-g,k} \leftarrow r_{x+1-g,k} + y$
**22**             $p_{x+1-g} \leftarrow p_{x+1-g} - y$
**23**             $r_{k,x+1-g} \leftarrow b - r_{x+1-g,k}$
**23**             $M \leftarrow M - y$
**24**         **for** $h \leftarrow g$ **downto** $1$
**25**             $A_{x+1-h} \leftarrow A_{x+1-h} - y$
**26 if** $M = 0$
**27**    **then for** $g \leftarrow 1$ **to** $k - 1$
**28**         **do** $r_{g,k} \leftarrow \max(r_{g,k}, 0)$
**29**             $r_{k,g} \leftarrow \min(r_{k,g}, b)$
**30**             **go to 41**

```
31 if A_x = 0
32    then for g ← k − 1 downto 1
33          do r_{g,k} ← max(r_{g,k}, 0)
34             for g ← k − 1 downto 1
35                do y ← max(a − r_{g,k}, 0)
36                   if M ≥ b − y
37                      then r_{k,g} ← y
38                           M ← M − (b − y)
39                      else r_{k,g} ← b − M
40                           M ← 0
41 for g ← 1 to 1
42    do q_g ← r_{g,k}
43       r_g ← r_{k,g}
44 return p, q, r
```

Let's demonstrate the work of MAIN and SCORESLICING by the reconstruction of the tournament whose point table is shown in Figure 1.

The basic idea is that MAIN slices (partitions) the points of $\mathcal{P}_6, \mathcal{P}_5, \ldots, \mathcal{P}_1$ by repeated calls of SCORESLICING.

The details are as follows. After assigning zeros to the elements of the main diagonal of $\mathcal{R}$ (in lines 01–03) MAIN calls SCORESLICING with $k = 6$. Then SCORESLICING computes the sequence of the additional points **A**, further the provisional last column and the provisional last row of $\mathcal{R}$ (lines 03-09). The results of the execution of lines 03–08 of SCORESET are represented in Figure 4.

| Player/Player | $\mathcal{P}_1$ | $\mathcal{P}_2$ | $\mathcal{P}_3$ | $\mathcal{P}_4$ | $\mathcal{P}_5$ | $\mathcal{P}_6$ | **p**$_6$ | **A** |
|---|---|---|---|---|---|---|---|---|
| $\mathcal{P}_1$ | — | | | | | 0 | 9 | 9 |
| $\mathcal{P}_2$ | | — | | | | 0 | 9 | 16 |
| $\mathcal{P}_3$ | | | — | | | 0 | 19 | 31 |
| $\mathcal{P}_4$ | | | | — | | 0 | 20 | 45 |
| $\mathcal{P}_5$ | | | | | — | 0 | 32 | 69 |
| $\mathcal{P}_6$ | 10 | 10 | 10 | 10 | 10 | — | 34 | 93 |

Figure 4: The results of lines 04–08 of SCORESLICING.

Line 09 yields the actual number of the missing points M, then in the lines 10–43 the sequences $\mathbf{p}_{k-1}$, **q**, and **r** are determined.

The steps of the reconstruction of the tournament are shown in Figure 5 in digital form. The second column of the figure contains the starting state of the reconstruction — the score sequence $\mathbf{p}_6 = (9, 9, 19, 20, 32, 34)$.

|  | $\mathbf{p}_6$ | $\mathbf{p}_6$ | $\mathbf{p}_6$ | $\mathbf{p}_6$ | $\mathbf{p}_5$ | $\mathbf{p}_5$ | $\mathbf{p}_5$ | $\mathbf{p}_4$ | $\mathbf{p}_3$ | $\mathbf{p}_2$ |
|---|---|---|---|---|---|---|---|---|---|---|
| $\mathcal{P}_1$ | 9 | 9 | 9 | 9 | 9 | 9 | 9 | 8* | 2* | 1* |
| $\mathcal{P}_2$ | 9 | 9 | 9 | 9 | 9 | 9 | 9 | 8* | 2* | 1* |
| $\mathcal{P}_3$ | 19 | 19 | 19 | 16* | 16 | 16 | 9* | 8* | 2* | – |
| $\mathcal{P}_4$ | 20 | 20 | 19* | 17* | 17 | 16* | 9* | 9 | – | – |
| $\mathcal{P}_5$ | 32 | 22* | 22 | 22 | 22 | 22 | 22 | – | – | – |
| $\mathcal{P}_6$ | 34 | 34 | 34 | 34 | – | – | – | – | – | – |

Figure 5: Steps of the reconstruction (stars denote changes).

The second column of Figure 6 contains the actual parameters k, x, $A_x$, M, f, d, m, and y.

| Parameter/k | 6 | 6 | 6 | 6 | 5* | 5 | 5 | 5 | 4* | 3* | 2* |
|---|---|---|---|---|---|---|---|---|---|---|---|
| x | 5 | 4* | 4 | 4 | 4 | 4 | 4 | 4 | 3* | 2* | – |
| $A_x$ | 69 | 59* | 58* | 53* | 39* | 38* | 24* | 12* | 18* | 2* | – |
| M | 16 | 6* | 5* | 0* | 18* | 17* | 3* | 0* | 21* | 18* | – |
| f | 1 | 1 | 2* | – | 1* | 2* | 4* | – | 3* | 2* | – |
| d | 12 | 1* | 10* | – | 1* | 9* | 9* | – | 8* | 2* | – |
| m | 10 | 1* | 3* | – | 1* | 7* | 1* | – | 6* | 0* | – |
| y | 10 | 1* | 2* | – | 1* | 7* | 1* | – | 6* | 0* | – |

Figure 6: Parameters of the reconstruction (stars denote changes).

$\mathcal{P}_5$ has $A_5 = 69 > 0$ additional points (computed in line 05) and $\mathcal{P}_6$ has $M = 16 > 0$ missing points (computed in line 9), therefore SCORESLICE executes lines 10–25. The algorithm determined in lines 11–13 that $\mathcal{P}_x = \mathcal{P}_5$ is the first player who can get from the missing points of $\mathcal{P}_6$. The frequency of players having $p_x$ points is $f = 1$ (computed in lines 14–16). The difference $p_{6,5} - p_{6,4} = 12$ (computed in line 17). At the moment we can slice at most $m = 10$ points per player (computed in line 18). Since $A_5$ is large enough we get $y = 10$ (computed in line 20), and decrease the number of points of $\mathcal{P}_5$ by $y = 10$ points (in line 21). Therefore the updated new values are $r_{5,6} = 10$, $r_{6,5} = 0$, $M = 6$ and $A_5 = 59$. The new score vector $\mathbf{p}_6 = (9, 9, 19, 20, 22^*, 34)$ is in the third column of Figure 5 (stars denote changes).

Since $M = 6 > 0$ and $A_5 = 59 > 0$, we use again lines 11–25 and since $r_{5,6} = 10$, we get a new, smaller value $x = 4$. $f$ remains 1, $d = 1$, $m = y = 1$, so $r_{4,6} = 1$, $p_4 = 19$, $r_{6,4} = 9$, $M = 5$, $A_4 = 58$. The new parameters are in the third column of Figure 6, the new score vector $\mathbf{p}_6 = (9, 9, 19, 19^*, 22, 34)$ appears in the fourth column of Figure 5.

Now $M = 5 > 0$ and $A_5 = 58 > 0$, so continuing with lines 10–25 $x$ remains 4 but the frequency is now $f = 2$, the difference $d = 10$, the small $M$ allows only $m = 3$ and $y = 3$ (see fourth column of Figure 6). So it follows $r_{3,6} = 3$, $p_3 = 16$, $r_{4,6} = 1 + 2 = 3$, $p_4 = 17$, $M = 0$, $A_5 = 53$, and $\mathbf{p}_6 = (9, 9, 16^*, 17^*, 22, 34)$ is shown in the fifth column of Figure 5. Since M decreased to zero, SCORESLICING continues in line 26 and executing line 44 returns to MAIN the sequences $\mathbf{p}_5 = (9, 9, 16^*, 17^*, 22)$, $\mathbf{q} = (10, 10, 7, 7, 0)$, and $\mathbf{r} = (0, 0, 3, 3, 10)$ shown in the sixth column of Figure 5, resp. in seventh line and seventh column of Figure 7.

| Player/Player | $\mathcal{P}_1$ | $\mathcal{P}_2$ | $\mathcal{P}_3$ | $\mathcal{P}_4$ | $\mathcal{P}_5$ | $\mathcal{P}_6$ | Score |
|---|---|---|---|---|---|---|---|
| $\mathcal{P}_1$ | — | 0 | 0 | 0 | 0 | **0** | 9 |
| $\mathcal{P}_2$ | 10 | — | 0 | 0 | 0 | **0** | 9 |
| $\mathcal{P}_3$ | 10 | 10 | — | 0 | 0 | **3** | 19 |
| $\mathcal{P}_4$ | 10 | 10 | 10 | — | 0 | **3** | 20 |
| $\mathcal{P}_5$ | 10 | 10 | 10 | 10 | — | **10** | 32 |
| $\mathcal{P}_6$ | **10** | **10** | **7** | **7** | **0** | — | 34 |

Figure 7: The partially reconstructed results of the matches of six players of the given tournament $\mathcal{T}_6(2, 10)$ after determining of the results of $\mathcal{P}_6$, where bold numbers denote final values.

After updating $\mathcal{R}$ MAIN calls SLICESCORING with the parameter $k = 5$.

The parameters determined in lines 11–16 are shown in the sixth column of Figure 6. Since $M = 18 > 0$ and $A_4 = 39 > 0$, the algorithm executes lines 11–25 and gets $r_{4,5} = 1$, $p_4 = 16$, $r_{5,4} = 9, M = 17$, and $A_4 = 38$. The new score vector $\mathbf{p}_4 = (9, 9, 16^*, 16, 22)$ is shown in the seventh column of Figure 6.

Since $M = 17 > 0$ and $A_4 = 38 > 0$, the algorithm in lines 11–16 computes the values shown in the seventh column of Figure 6 and then in lines 18–23 gets $r_{3,6} = 1 + 7 = 8$, $p_3 = 9$, $r_{6,3} = 2$, $r_{4,6} = 0 + 7 = 7$, $p_4 = 9$, $r_{6,4} = 3$, $M = 3$, and $A_4 = 24$. The new score vector $\mathbf{p}_5 = (9^*, 9^*, 9, 9, 22)$ is shown in the tenth column of Figure 6.

Now $M = 3 > 0$ and $A_4 = 24 > 0$, therefore the algorithm continues in line 11 and gets the parameter values contained in the eighth column of Figure 6. These values imply in lines 18–25 $r_{1,5} = 1$, $p_1 = 8$, $r_{2,5} = 1$, $p_2 = 8$, $r_{3,5} = 1$, $p_3 = 8$, $\mathbf{p}_4 = (8, 8, 8, 9)$ and $M = 0$. Since $M = 0$, the algorithm continues in line 26 and in lines 26–30 gets $\mathbf{q} = (1, 1, 8, 8)$ and $\mathbf{r} = (9, 9, 2, 2)$. SCORESLICING returns these vectors to MAIN and it finishes the filling of the sixth line and sixth column of $\mathbf{R}$. The resulted $\mathbf{R}$ is shown in Figure 8.

MAIN continues by calling SCORESLICING for $k = 4$. Since $M = 21 > 0$ and $A_3 = 18 > 0$, the algorithm gets in lines 11–16 the parameters shown in the ninth column of Figure 6. Line 20 results $y = 6$ due to the small amount of additional points of $\mathcal{P}_3$. So we get $r_{1,4} = 6$, $p_1 = 2$, $r_{4,1} = 4$, $r_{2,4} = 6$, $p_2 = 2$, $r_{4,2} = 4$, $r_{3,4} = 6$, $p_3 = 2$, $r_{4,3} = 4$, $M = 0$, then $\mathbf{p} = (2, 2, 2)$, $\mathbf{q} = (6, 6, 6)$ and $\mathbf{r} = (3, 3, 3)$. Using the returned vectors MAIN fills the fifth row and the fifth column of $\mathbf{R}$ as Figure 9 shows.

| Player/Player | $\mathcal{P}_1$ | $\mathcal{P}_2$ | $\mathcal{P}_3$ | $\mathcal{P}_4$ | $\mathcal{P}_5$ | $\mathcal{P}_6$ | Score |
|---|---|---|---|---|---|---|---|
| $\mathcal{P}_1$ | — | 0 | 0 | 0 | **1** | **0** | 9 |
| $\mathcal{P}_2$ | 10 | — | 0 | 0 | **1** | **0** | 9 |
| $\mathcal{P}_3$ | 10 | 10 | — | 0 | **8** | **3** | 19 |
| $\mathcal{P}_4$ | 10 | 10 | 10 | — | **8** | **3** | 20 |
| $\mathcal{P}_5$ | **9** | **9** | **2** | **2** | — | **10** | 32 |
| $\mathcal{P}_6$ | **10** | **10** | **7** | **7** | **0** | — | 34 |

Figure 8: The partially reconstructed results of the matches of six players of the given tournament $\mathcal{T}_6(2, 10)$ after determining of the results of $\mathcal{P}_5$, where bold numbers denote final values.

MAIN continues by calling SCORESLICING for $k = 3$. Since now $M = 18 > 0$, and $A_2 = 2 > 0$, the algorithm gets in lines 11–16 the parameters shown in the tenth column of Figure 6. So lines 18-25 give the results $r_{1,3} = 1, p_1 = 1$, $r_{2,3} = 1$, $p_2 = 1$, and $M = 0$. Then we get in lines 26–29 that $\mathbf{q} = (1, 1)$ and $\mathbf{r} = (1, 1)$. Using the returned vectors MAIN fills the fifth row and the fifth column of $\mathbf{R}$, then in lines 10–11 determines $r_{1,2}$ and $r_{2,1}$.

Figure 10 shows the point table of the reconstructed tournament.

Figure 11 shows the rounds of the reconstruction in graphical form.

| Player/Player | $\mathcal{P}_1$ | $\mathcal{P}_2$ | $\mathcal{P}_3$ | $\mathcal{P}_4$ | $\mathcal{P}_5$ | $\mathcal{P}_6$ | Score |
|---|---|---|---|---|---|---|---|
| $\mathcal{P}_1$ | — | 0 | 0 | **6** | **1** | **0** | 9 |
| $\mathcal{P}_2$ | 10 | — | 0 | **6** | **1** | **0** | 9 |
| $\mathcal{P}_3$ | 10 | 10 | — | **6** | **8** | **3** | 19 |
| $\mathcal{P}_4$ | **3** | **3** | **3** | — | **8** | **3** | 20 |
| $\mathcal{P}_5$ | **9** | **9** | **2** | **2** | — | **10** | 32 |
| $\mathcal{P}_6$ | **10** | **10** | **7** | **7** | **0** | — | 34 |

Figure 9: The partially reconstructed results of the matches of six players of the given tournament $\mathcal{T}_6(2, 10)$ after determining of the results of $\mathcal{P}_4$, where bold numbers denote final values.

| Player/Player | $\mathcal{P}_1$ | $\mathcal{P}_2$ | $\mathcal{P}_3$ | $\mathcal{P}_4$ | $\mathcal{P}_5$ | $\mathcal{P}_6$ | Score |
|---|---|---|---|---|---|---|---|
| $\mathcal{P}_1$ | — | 1 | 1 | 6 | 1 | 0 | 9 |
| $\mathcal{P}_2$ | 1 | — | 1 | 6 | 1 | 0 | 9 |
| $\mathcal{P}_3$ | 1 | 1 | — | 6 | 8 | 3 | 19 |
| $\mathcal{P}_4$ | 3 | 3 | 3 | — | 8 | 3 | 20 |
| $\mathcal{P}_5$ | 9 | 9 | 2 | 2 | — | 10 | 32 |
| $\mathcal{P}_6$ | 10 | 10 | 7 | 7 | 0 | — | 34 |

Figure 10: The fully reconstructed results of the matches of players of the given tournament $\mathcal{T}_6(2, 10)$.

### 3.3.1 Complexity analysis of ScoreSlicing and Main

The running time of this algorithm equals to $O(bn^3)$, since the sum of the missing points $M_k$ is $O(bk^2)$, and the sum of the additional points $A_k$ is $O(bk^2)$, and the sum of the scores $s_i$ is $O(bn^2)$, and the processing of a missing point, of an additional point and also of a win point requires $O(n)$ steps.

The memory requirement of SCORESLICING equals to $\Theta(n^2)$.

The running time of lines 01–03 of MAIN is $\Theta(n)$. In lines 04–09 algorithm SCORESLICING is executed $\Theta(n)$ times, so the running time of MAIN depends on the running time of SCORESLICING and is $O(bn^3)$.

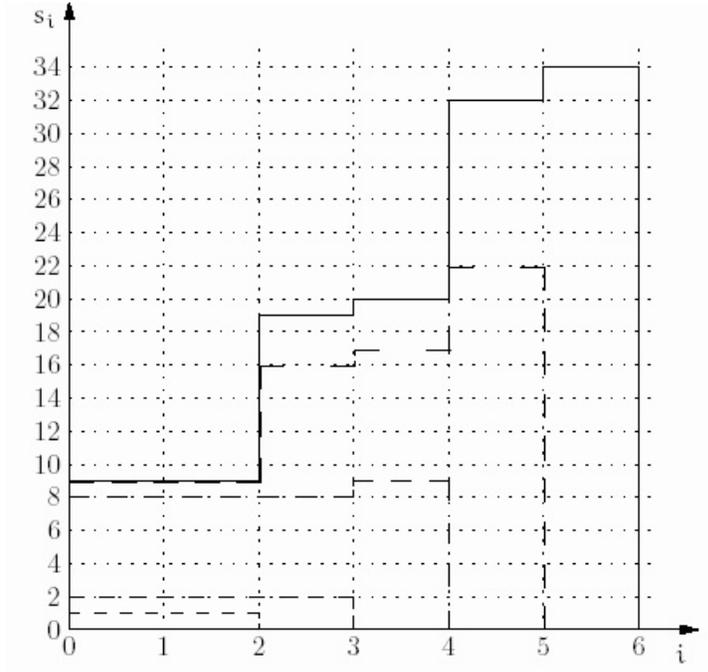

Figure 11: The staircase functions of the score sequences $\mathbf{p}_6 = (9, 9, 19, 20, 32, 34)$, $\mathbf{p}_5 = (9, 9, 16, 17, 22)$, $\mathbf{p}_4 = (8, 8, 8, 9)$, $\mathbf{p}_3 = (2, 2, 2)$, and $\mathbf{p}_2 = (1, 1)$.

## 4 Necessary and sufficient condition for $(a, b)$-tournaments

**Theorem 3** *A sequence* $(s_1, s_2, \ldots, s_n)$ *satisfying* $0 \leq s_1 \leq s_2 \leq \cdots \leq s_n$ *is the score sequence of some tournament* $\mathcal{T}_n(a, b)$ *if and only if*

$$aB_k \leq \sum_{i=1}^{k} s_i \leq bB_n - L_k - (n-k)s_i \ (1 \leq k \leq n).$$

**Proof.** Lemma 3 implies the necessity of these inequalities.

The sufficiency of these inequalities can be shown by induction based on the correctness of the reconstruction algorithm.

If $n = 2$, then $a \leq s_1 + s_2 \leq b$ due to 6 and then the scores $r_{1,2} \leftarrow \lfloor S_2/2 \rfloor$ and $r_{2,1} \leftarrow \lceil S_2/2 \rceil$ received by lines 10 and 11 of MAIN are correct values.

Let now $n > 2$. It is sufficient to show that SCORESLICING reduces the input problem of size $n$ to the reconstruction of the scores of $n-1$ players.

$A_k = S_k - aB_k \leq bB_k - aB_k$ and $M = b(n-1)$ imply $\min(A_k, M) \leq \min((b-a)B_k, b(n-1)) \leq bn(n-1)/2$. This minimum decreases at least by 1 in each execution of the while cycle in lines 23 and 25 – or at least one of $M$ and $A_k$ becomes to zero (if $f = 1$, then $A_k > 0$ due to line 10, and if $f \geq 2$, then $A_{x+1-g} > 0$, since otherwise $A_{x-g} < 0$, what is impossible) and SCORESLICING ends quickly in lines 26–30 or in lines 31-40.

The inequality (8) is guaranteed by lines 18, 20, and 35.
The inequality (9) is guaranteed by lines 18 and 20.
The inequality (10) is guarantedd by line 20.
The inequality (11) is guaranteed by line 19–23. ∎

**Acknowledgement.** The author thanks Bence Sári (Eötvös Loránd University of Budapest) and Tibor Liska (Computer and Automation Research Institute of HAS) for the computer experiments, András Gyárfás (Computer and Automation Research Institution of HAS) and Béla Vizvári (Eötvös Loránd University of Budapest) for their interest and useful comments.